\documentclass[superscriptaddress,prl,twocolumn,amsmath,amssymb]{revtex4}
\pdfoutput=1
\usepackage{graphicx}
\usepackage{eurosym}
\usepackage{dcolumn}
\usepackage{bm}
\usepackage{subfigure}
\usepackage[final]{pdfpages}
\begin{document}

\newcommand{\tr}{\operatorname{Tr}}
\newcommand{\ket}[1]{\left | #1 \right \rangle}
\newcommand{\bra}[1]{\left \langle #1 \right |}
\newcommand{\proj}[1]{\ket{#1}\!\!\bra{#1}}

\title{Quantum walks of correlated particles}

\author{Alberto Peruzzo}
\author{Mirko Lobino}
\author{Jonathan C. F. Matthews}
\affiliation{Centre for Quantum Photonics, H. H. Wills Physics Laboratory \& Department of Electrical and Electronic Engineering, University of Bristol, Merchant Venturers Building, Woodland Road, Bristol, BS8 1UB, UK}
\author{Nobuyuki Matsuda}
\affiliation{\mbox{NTT Basic Research Laboratories, NTT Corporation 3-1 Morinosato Wakamiya, Atsugi, Kanagawa 243-0198, Japan}} 
\affiliation{\mbox{Research Institute of Electrical Communication, Tohoku University, Sendai, Miyagi 980-8577, Japan}}
\author{Alberto Politi}
\author{Konstantinos Poulios}
\author{Xiao-Qi Zhou}
\affiliation{Centre for Quantum Photonics, H. H. Wills Physics Laboratory \& Department of Electrical and Electronic Engineering, University of Bristol, Merchant Venturers Building, Woodland Road, Bristol, BS8 1UB, UK}
\author{Yoav Lahini}
\affiliation{Department of Physics of Complex Systems, The Weizmann Institute of Science, Rehovot, Israel}
\author{Nur Ismail}
\author{Kerstin W\"{o}rhoff}
\affiliation{\mbox{Integrated Optical Microsystems Group, MESA+ Institute for Nanotechnology, University of Twente, Enschede, The Netherlands}}
\author{Yaron Bromberg}
\author{Yaron Silberberg}
\affiliation{Department of Physics of Complex Systems, The Weizmann Institute of Science, Rehovot, Israel}
\author{Mark G. Thompson}
\author{Jeremy L. O'Brien}
\email{Jeremy.OBrien@Bristol.ac.uk}
\affiliation{Centre for Quantum Photonics, H. H. Wills Physics Laboratory \& Department of Electrical and Electronic Engineering, University of Bristol, Merchant Venturers Building, Woodland Road, Bristol, BS8 1UB, UK}

\begin{abstract}

Quantum walks of correlated particles offer the possibility to study large-scale quantum interference, simulate biological, chemical and physical systems, and a route to universal quantum computation. Here we demonstrate quantum walks of two identical photons in an array of 21 continuously evanescently-coupled waveguides in a SiO$_x$N$_y$ chip. We observe quantum correlations, violating a classical limit by 76 standard deviations, and find that they depend critically on the input state of the quantum walk. These results open the way to a powerful approach to quantum walks using correlated particles to encode information in an exponentially larger state space.

\end{abstract}

\maketitle

\begin{figure}[b!]
    \centering
    \includegraphics[width = 8cm]{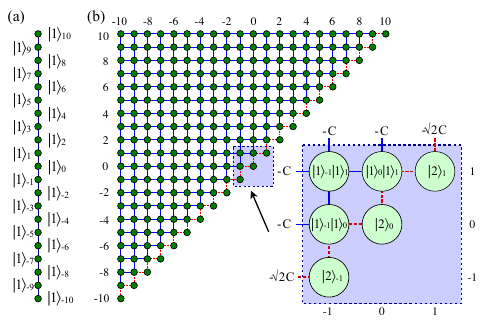}
    \vspace{-0.25cm}
\caption{\footnotesize{Quantum walks with one and two indistinguishable photons described by a Hamiltonian of coupled harmonic oscillators (Eq.~\ref{HarmonicOscillators}). (a). The linear array of vertices representing the state space of one photon populating $N=21$ waveguides in a coupled one-dimensional array. Site potentials for each vertex are $-\beta$ while hopping amplitudes between nearest neighbours are all equal to $-C$. (b). The two-dimensional lattice of vertices that represent the state space of two photons populating $N=21$ waveguides in a coupled one-dimensional waveguide array. (inset) Enlarged portion of the lattice displays the vertex representation of the two photon basis states. Site potentials are all equal to $-2\beta$, while hopping amplitudes between adjacent vertices are either $-C$ or $-\sqrt{2} C$ as labeled.}}
\label{2Dlattice}
\end{figure}

\begin{figure}[t]
\centering 
	\includegraphics[width=\columnwidth]{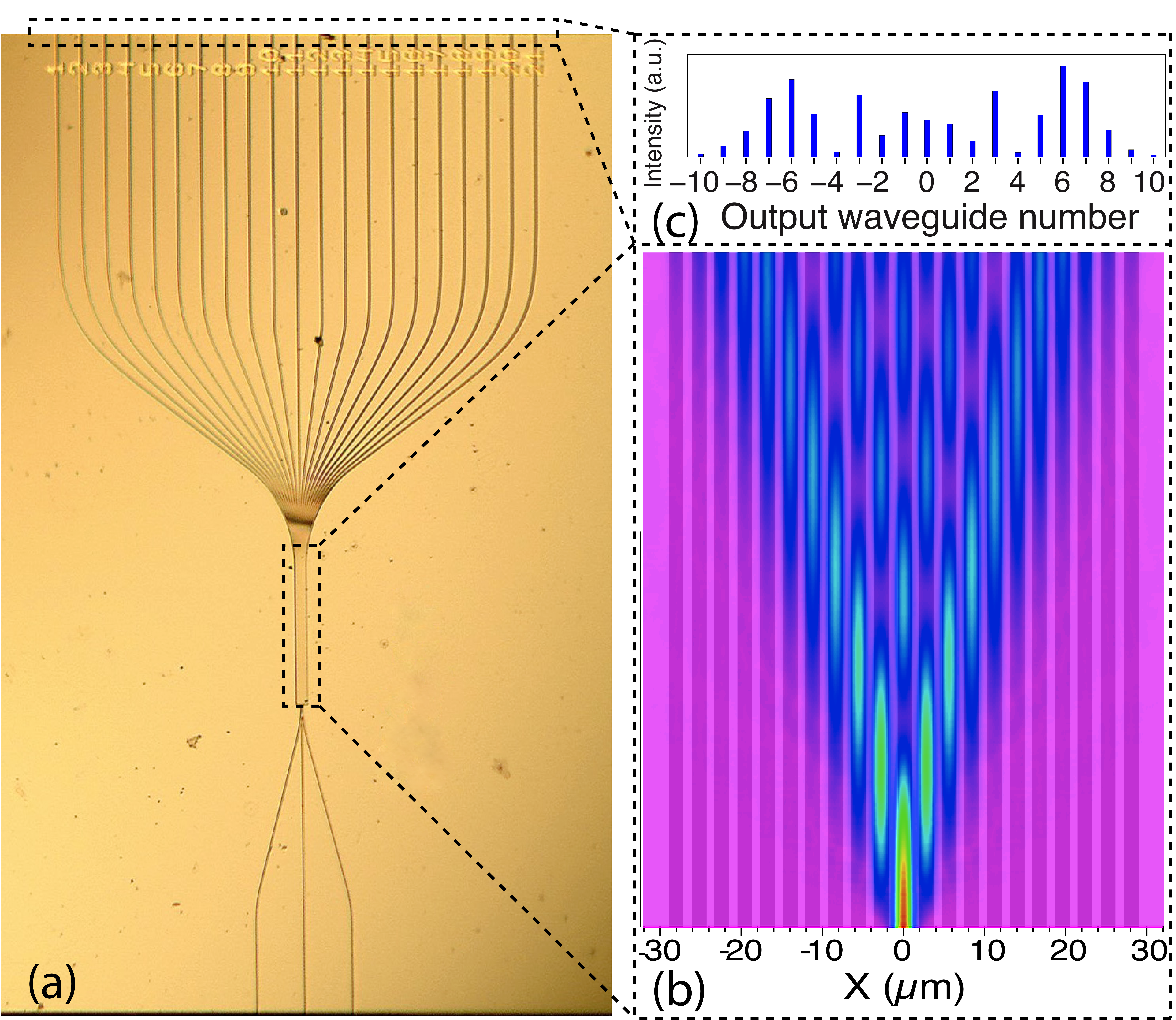}
	\vspace{-0.25cm}
\caption{A continuously coupled waveguide array for realising correlated photon quantum walks. (a) An optical micrograph of a 21 waveguide array showing the three input waveguides, initially separated by 250$\mu$m bending into the 700$\mu$m long coupling region. All 21 outputs bend out to 125$\mu$m spacing. (b) Simulation of single photon propagation in the array. (c) Output pattern of 810nm laser light propagating through the waveguide array.} 
\label{fig_chipPicture} 
\vspace{-0.5cm}
\end{figure}

With origins dating back to observations by Lucretius in 60BC and Brown in the 1800's, random walks are a powerful tool used in a broad range of fields from genetics to economics \cite{MotwaniRandom}. The quantum mechanical analogue---quantum walks \cite{ah-pra-48-1687,fa-pra-58-915}---corresponds to the tunnelling of quantum particles into several possible sites, generating large coherent superposition states and allowing massive parallelism in exploring multiple trajectories through a given connected graph (\emph{eg.} Fig.~\ref{2Dlattice}). This quantum state evolution is a reversible (unitary) process and so requires low noise (decoherence) systems for observation. In contrast to the diffusive behaviour of (classical) random walks, which tend towards a steady state, the wave function in a quantum walk propagates ballistically (Fig.~\ref{fig_chipPicture}(c)). These features are at the heart of new algorithms for database-search \cite{ch-pra-70-022314}, random graph navigation, models for quantum communication using spin chains \cite{bo-prl-91-207901}, universal quantum computation \cite{ch-prl-102-180501} and quantum simulation \cite{mo-jcp-129-174106}.

Quantum walks have been demonstrated using nuclear magnetic resonance \cite{ry-pra-72-062317,du-pra-77-042316}, phase \cite{sc-prl-103-090504,za-prl-104-100503} and position \cite{ka-sci-325-174} space of trapped ions, the frequency space of an optical resonator \cite{bo-pra-61-013410}, single photons in bulk \cite{do-josab-22-499} 
 and fibre \cite{sc-prl-104-050502} optics and the scattering of light in coupled waveguide arrays \cite{pe-prl-100-170506}. 
However, to date, all realisations have been limited to single particle quantum walks, which have an exact mapping to classical wave phenomena \cite{kn-pra-68-020301}, and therefore cannot provide any advantage from quantum effects  (note that the quantum walk with two trapped ions \cite{za-prl-104-100503} encodes in the centre of mass mode and is therefore effectively a single particle quantum walk on a line). Indeed single particle quantum walks have been observed using classical light \cite{pe-prl-100-170506,ch-nat-424-817}. 
In contrast, for quantum walks of more than one indistinguishable particle, classical theory no longer provides a sufficient description---quantum theory predicts that probability amplitudes interfere leading to distinctly non-classical correlations \cite{om-pra-74-042304,pa-pra-75-032351}. This quantum behaviour gives rise to a computational advantage in quantum walks of two identical particles, which can be used to solve the graph isomorphism problem for example \cite{ga-pra-81-052313}. 
The major challenge associated with realising quantum walks of correlated particles is the need for a low decoherence system that preserves their non-classical features.

The intrinsically low decoherence properties and easy manipulation of single photons make them ideal for observing quantum mechanical behaviour and for quantum technologies \cite{ob-natphot-3-687}; and the effectiveness of arrays of continuously coupled waveguides for bright classical light has been demonstrated \cite{pe-prl-100-170506,ch-nat-424-817}. However, quantum walks with correlated photons in such structures requires a means to measure two-photon correlations across the waveguide array. The spacing between waveguides in an array required for evanescent coupling (of order several $\mu$m) is significantly smaller than the minimum spacing of optical fibre arrays (127 $\mu$m) typically used to couple to single photon detectors. Previous quantum optical waveguide circuits in a silica-on-silicon architecture \cite{ob-natphot-3-687} promise to avoid decoherence effects \cite{ke-mscs-17-1169} present in other experimental realisations, interferometric stability and near-perfect mode overlap \cite{ob-natphot-3-687}, which is problematic in large scale bulk optical realisations (an arbitrary $N$ mode multi-port would require a network of $O(N^2)$ beams splitters \cite{re-prl-73-58}). 
However,  the low refractive index contrast ($\Delta = (n_{core}^2 - n_{cladding}^2)/2 n_{core}^2 \approx0.5\%$) in this architecture results in a large minimum bend radius ($<0.1$dB loss at 800 nm) of $\approx15$ mm, making it unsuitable for coupling into and out of large array quantum walk devices. 

We have overcome these technical challenges by fabricating waveguide arrays in SiO$_x$N$_y$ (silicon oxynitride), a material that enables a much higher refractive index contrast than silica-on-silicon (the refractive index is determined by $x$ and $y$), resulting in more compact devices (Fig.~\ref{fig_chipPicture}(a)) and a practical means to realise large coupled waveguide arrays that can be coupled to optical fibres. The device shown in Fig.~\ref{fig_chipPicture}(a) is a 5 mm long silicon chip with SiO$_x$N$_y$ waveguides with high refractive index contrast $\Delta = 4.4 \%$. The minimum bend radius for this index contrast is 600 $\mu\textrm{m}$ which enables much more rapid spreading of the waveguides from the evanescent coupling region where waveguides are pitched at 2.8$\mu$m to a pitch suitable for optical fibre (250$\mu$m and 125$\mu$m for photon injection and photon collection respectively, see Appendix).

\begin{figure*}[t!]
\includegraphics[width=\textwidth]{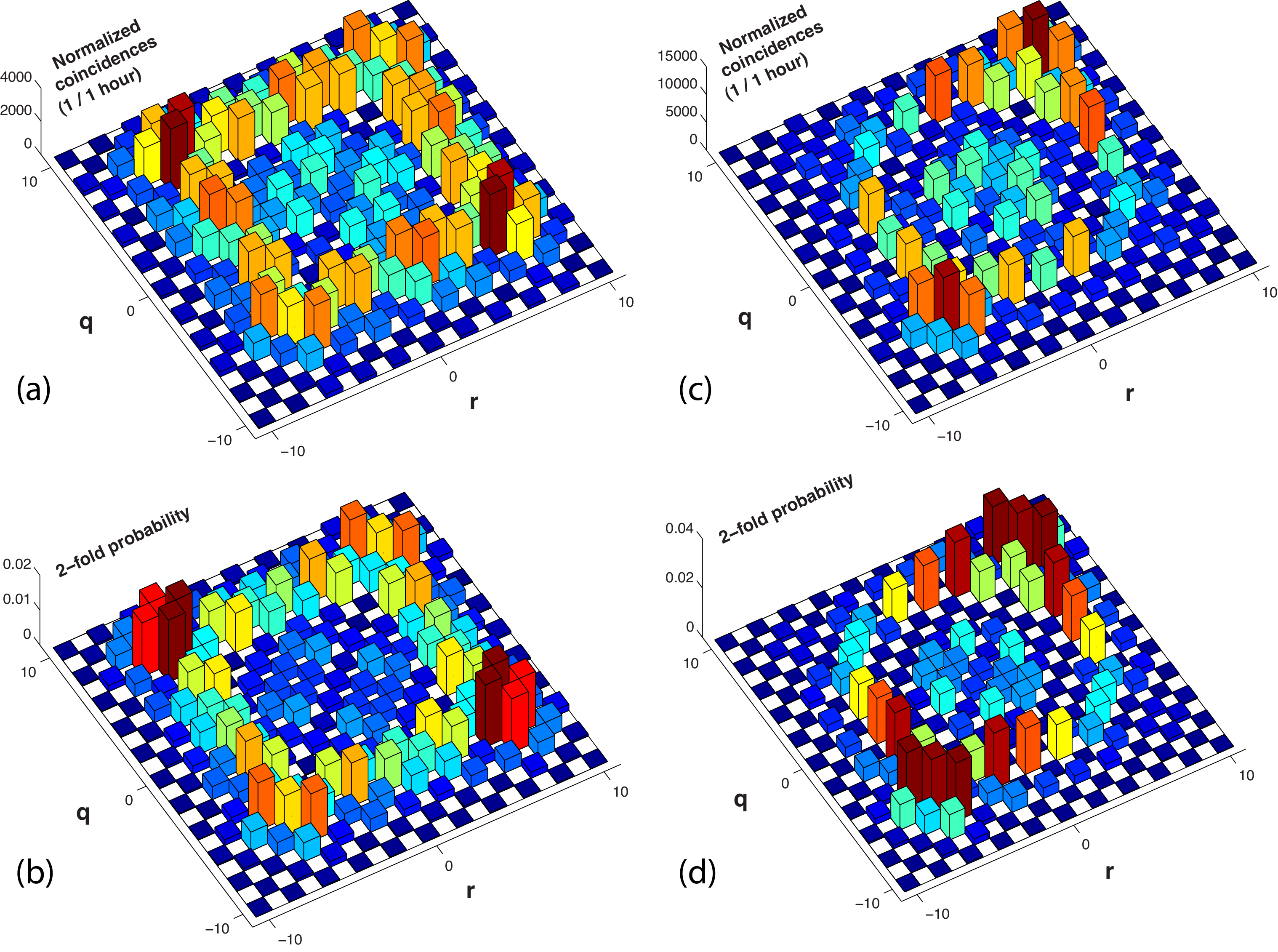}
\caption{Measured and simulated correlations in waveguide arrays when two photons are coupled to waveguides 0 and 1: $a^\dagger_0 a^\dagger_1\ket{0}$ (a,b) for input photons separated with temporal delay longer than their coherence length and (c,d) for photons arriving simultaneously in the array. All resulting measurements are corrected for coupling fluctuations using simultaneously detected single photon signal and as well relative detector efficiency; the integration time was 1 hour. The outcome of two photons populating one waveguide was detected using non-deterministic photon number resolving detection using an optical fibre splitter. }
\label{011results}
\vspace{-0.5cm}
\end{figure*}

\begin{figure*}[t!]
\includegraphics[width=\textwidth]{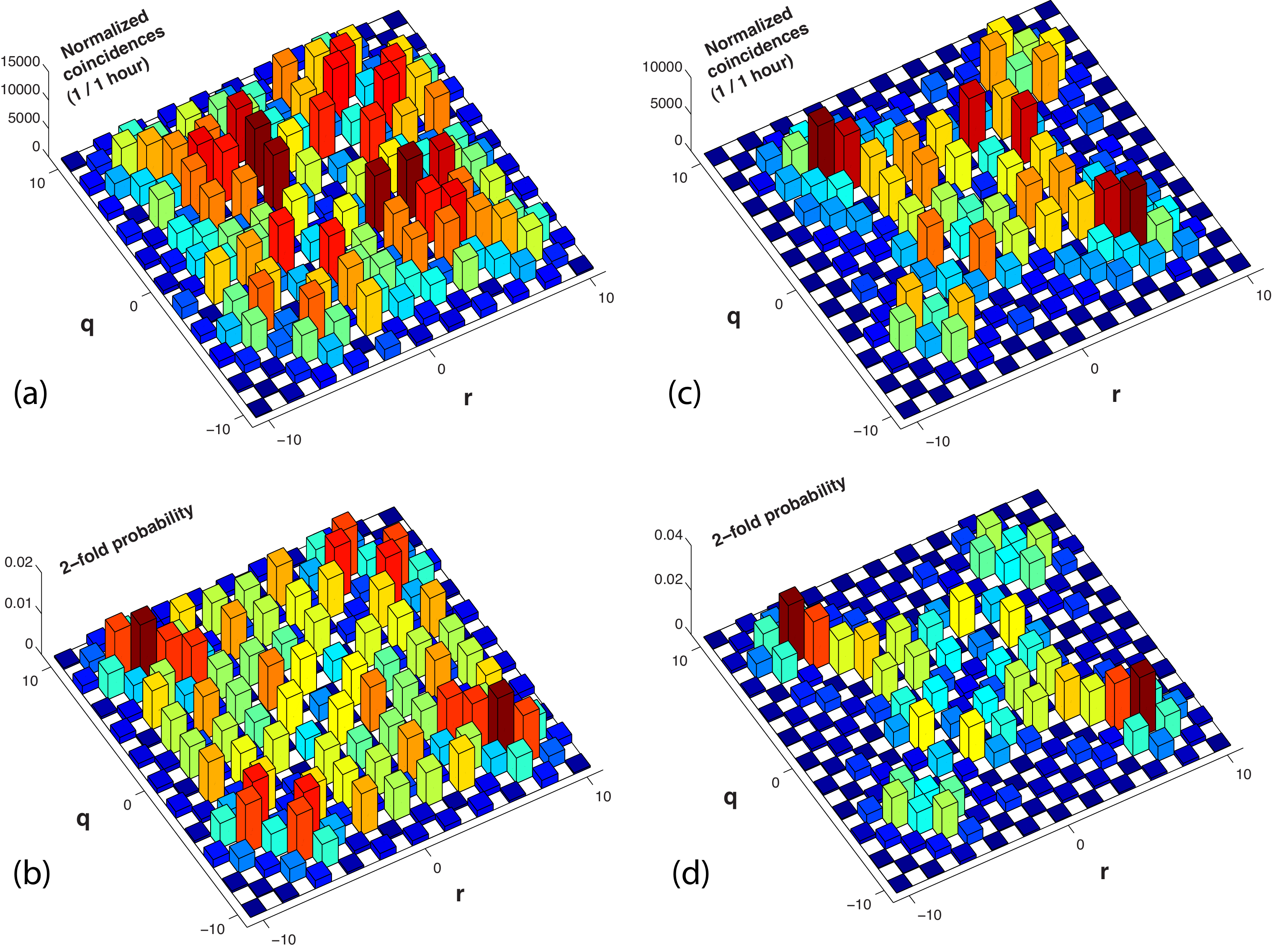}
\caption{Measured and simulated correlations in waveguide arrays when two photons are coupled to waveguides -1 and 1: $a^\dagger_{-1} a^\dagger_1\ket{0}$ (a,b) for input photons separated with temporal delay longer than their coherence length and (c,d) for the photons arriving simultaneously in the array. All measurements are corrected for coupling fluctuations using the simultaneously detected single photon count rate and relative detector efficiency; the integration time was 1 hour. The outcome of two photons populating one waveguide was detected using non-deterministic photon number resolving detection using an optical fibre splitter.}
\vspace{-0.5cm}
\label{101results}
\end{figure*}

Photon pairs were generated via type-I SPDC in a 2 mm thick $\chi^2$ nonlinear bismuth borate $\textrm{BiB}_3\textrm{O}_6$ (BiBO) crystal, pumped with 40 mW of 402 nm light from a CW diode laser. With an opening angle of $3^\circ$, pairs of degenerate $\lambda=804$ nm photons were filtered by 2 nm FWHM interference filters, and focused onto two inputs of an array of polarization maintaining fibres which were butt-coupled to the waveguide chip. At the output of the chip, an array of multi mode fibres guide the output to 12 single photon counting modules connected to photon counting logic based on three FPGA boards used to measure matrices of two-photon correlations between the outputs of the array (Figs.~\ref{011results} and \ref{101results}). Quantum interference (degree of indistinguishability) was controlled by relative temporal delay between the pair of photons, using an automated linear actuator, and characterised with a standard Hong-Ou-Mandel experiment \cite{ho-prl-59-2044}. The pitch of the waveguide outputs is half of that of the collecting fibre arrays allowing 121 of the $231$ possible two-photon correlations to be measured at the output. 

Photons propagating through the coupled waveguide array shown in Fig.~2(a) are modelled assuming nearest neighbour interaction with the Hamiltonian for coupled oscillators \cite{es-pr-175-286}, setting $\hbar=1$:
\vspace{-0.15cm}
\begin{eqnarray}
\hat{H} = \sum_{j=1}^N\left[\beta_j a_j^\dagger a_j + C_{j,j-1} a^\dagger_{j-1} a_j +C_{j,j+1} a^\dagger_{j+1} a_j \right],
\label{HarmonicOscillators}
\end{eqnarray}
\vspace{-0.25cm}

\noindent where the creation and annihilation operators  $a^\dagger_j$ and $a_j$ obey Bose-Einstein statistics and act on waveguide $j$. In our devices the waveguide propagation constant $\beta_j=\beta$ and coupling constant between adjacent waveguides $C_{j,j\pm1}=C=C$ are designed to be uniform for all $j$. Through choice of operator $\hat{A}$ the Heisenberg equation of motion $i d \hat{A}/d z=[\hat{A},\hat{H}]$ models the dynamics of photons propagating along distance $z$ of the array. For example, injecting single photons in waveguide $k$ is described using $\hat{A} = a^\dagger_k$, yielding \cite{br-prl-102-253904}
\vspace{-0.15cm}
\begin{eqnarray}
i\frac{\textrm{d} a^\dagger_k}{\textrm{d} z} = - \beta a_k^\dagger  - C a^\dagger_{k-1} - C a^\dagger_{k+1}.
\label{hbergsingle}
\end{eqnarray}
\vspace{-0.25cm}

\noindent Acting Eq.~\ref{hbergsingle} on the vacuum $\ket{0}$, and inspecting in the Schr\"{o}dinger picture $i\textrm{d}\ket{\psi_1}/\textrm{d}z=H^{(1)}\ket{\psi_1}$, yields $H^{(1)}$. From this Hamiltonian, an adjacency matrix $M^{(1)}_{j,k} = \mbox{}_j\bra{1} H^{(1)} \ket{1}_k$ of the graph in Fig.~\ref{2Dlattice}(a) is constructed using the single photon Fock basis $\{\ket{1}_j\}$ for representation.
A quantum walk on this graph evolves over length $z$ according to the unitary transform $U^{(1)}=\textrm{exp}[-i H^{(1)} z]$ \cite{fa-pra-58-915,br-prl-102-253904}, the exact dynamics of which can be observed by injecting bright light into the waveguide array \cite{pe-prl-100-170506}. We used this approach to calibrate our device and the coupling efficiencies by launching horizontally polarised 810 nm laser light into the input of the central waveguide, waveguide 0. We measured the interference pattern shown in Fig.~\ref{fig_chipPicture}(c) which corresponds to the probability distribution for single photons detected at the 21 output waveguides (numbered -10 on the left, through 0, to 10 on the right). Before the waveguides reach their minimum separation at the central coupling region, they are significantly coupled in the spreading regions on both sides. This coupling provides an effective coupling length of 82 $\mu$m.  
We determined this length by comparing the output interference pattern  (not shown) for an array of length 350 $\mu$m length. 
From these data the coupling constant $C=5\mbox{ mm}^{-1}$ was determined by running the simulation and minimising the square of the errors.

The injection of two indistinguishable photons is modelled with the operator $\hat{A} = a^\dagger_j a^\dagger_k$ yielding: 
\vspace{-0.15cm}
\begin{eqnarray}
i\frac{\textrm{d} a^\dagger_j a^\dagger_k}{\textrm{d} z}\!\!=\!\!
-\!2\beta a^\dagger_j a^\dagger_k\!\! -\!C\!\! \left[a_j^\dagger a^\dagger_{k-1}\!\! +\! a_j^\dagger a^\dagger_{k+1}\!\!
+\! a_k^\dagger a^\dagger_{j-1}\!\!+\! a_k^\dagger a^\dagger_{j+1}\!\right]\label{hbergdouble}
\end{eqnarray}
\vspace{-0.25cm}

\noindent From Eq.~\ref{hbergdouble}, $H^{(2)}$ acting on the two photon Fock space is extracted as a matrix represented in the two-photon Fock basis $\{\ket{1}_j\ket{1}_k,\ket{2}_l\}$ which is equal to the adjacency matrix for the graph given in Fig.~\ref{2Dlattice}(b). Evolution of the two-photon state through the device therefore simulates a single particle quantum walk on the graph in Fig.~\ref{2Dlattice}(b) with $O(N^2)$ vertices and unitary transform  $U^{(2)} = \textrm{exp}[-i H^{(2)}z]$. 
In general, linear increase in the number of photons input into the coupled array results in exponential growth of the Hilbert space and of the corresponding graph on which a single particle quantum walk is emulated. However, only when indistinguishable photons are injected in the device can the output state be non-seperable, exhibiting non-classical correlated behaviour ({see Appendix}).
Note that the two photon unitary evolution can also be computed from the product of single photon mode transformations \cite{br-prl-102-253904}.
 
The measured correlation matrices $\Gamma_{q,r}$ (defined as probability to detect a two photon coincidence across waveguides $q,r$) \cite{ma-apb-60-s111,br-prl-102-253904} for injecting two single photons into the central neighbouring waveguides 0 and 1 (\emph{i.e.}  $a^\dagger_0 a^\dagger_1\ket{0}$) are plotted in Fig. \ref{011results}(a) for photons made distinguishable using temporal delay (not overlapped) and in Fig. \ref{011results}(c) for pairs of indistinguishable (overlapped) photons. The overlap of these measured distributions with ideal simulations (plotted in Fig.~\ref{011results}(b) and \ref{011results}(d)) are $S=0.980\pm 0.001$ and $S=0.934\pm 0.001$, respectively, where $S$ is the similarity between two probability distributions $\Gamma$ and $\Gamma'$ defined by $S= (\sum _{i,j}\sqrt{\Gamma_{i,j} \Gamma'_{i,j}})^2/{\sum _{i,j}\Gamma_{i,j} \sum _{i,j}\Gamma'_{i,j}}$, which is a generalisation of the average fidelity based on the (classical) fidelity between probability distributions. 
The lower $S$ in the overlapped case is attributed to imperfect quantum interference. These results clearly display a generalised bunching behaviour (tending to both travel to one side of the array or the other), characteristic of quantum interference: the vanishing of the two off-diagonal lobes is a result of destructive interference of quantum amplitudes resulting from repeated $\pi/2$ phase shifts in the photon tunnelling between neighbouring waveguides.

\begin{figure*}[t!]
\vspace{-0.25cm}
\includegraphics[width=\textwidth]{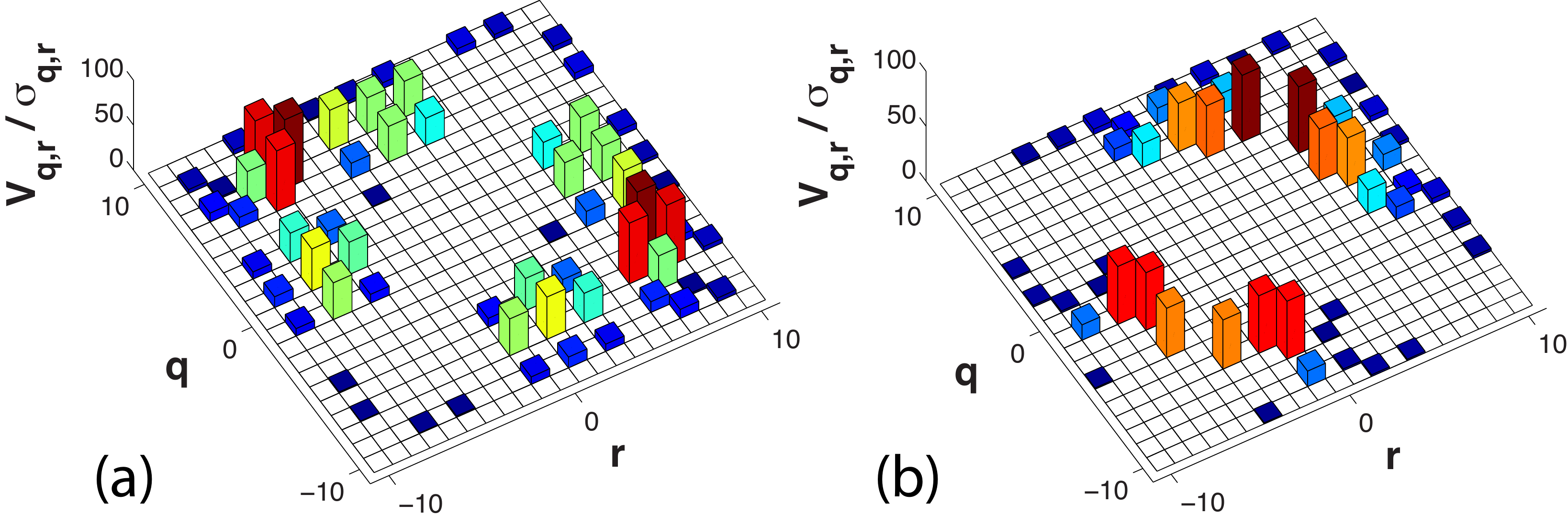}
\caption{Violating the classical limit. Number of standard deviations $\sigma$ of the violation of inequality Eq.~\ref{eq:6} for injecting indistinguishable photons into the inputs 0, 1 (a) and -1, 1 (b) and measuring correlations at waveguide outputs labelled $q$ and~$r$.}
\label{Number of s violated}
\vspace{-0.5cm}
\end{figure*}

Distinctly different behaviour is observed on injecting two photons in two waveguides with one waveguide separating them. The measured correlation matrices for injecting photons into waveguides $-$1 and 1, and the vacuum in waveguide 0 (\emph{i.e.}  $a^\dagger_{-1} a^\dagger_1\ket{0}$) in the centre of the array are plotted in Fig \ref{101results}. The similarities with the ideal simulation are $S=0.970\pm 0.002$ and $S=0.903\pm0.002$ for the delayed and overlapped photons, respectively. 
In this case, instead of bunching, when the two photons are indistinguishable they generate a pattern where the main feature is the vanishing of probability to simultaneously detect one photon in the centre of the array and a one  at the limit of ballistic propagation (for example in waveguides 0 and 7 in Fig \ref{101results}(c),(d)).

Detecting two-fold coincidences of two indistinguishable photons leads to non-classical correlations across pairs of waveguide in the array \cite{br-prl-102-253904}. 
The correlation function after length $z$ for two photons populating waveguides $q$ and $r$ is given \cite{ma-apb-60-s111,br-prl-102-253904} by:
\begin{eqnarray}
\Gamma_{q,r}(z) =\frac{1}{1+\delta_{q,r}}\left|U^{(1)}_{q,q'}U^{(1)}_{r,r'}+U^{(1)}_{q,r'}U^{(1)}_{r,q'}\right|^{2}
\label{eq:4}
\end{eqnarray}
For classical light, including random phase fluctuations that mimic certain properties of quantum light, diagonal correlations $\Gamma_{q,q}$ are related to correlations in the off-diagonal lobes $\Gamma_{q,r}$, $q\neq r$ according to the inequality \cite{br-prl-102-253904}:
\begin{eqnarray}
V_{q,r}=\Gamma^{(Cl)}_{q,r} - \frac{{2}}{3} \sqrt{\Gamma^{(Cl)}_{q,q}\,\,\Gamma^{(Cl)}_{r,r}} > 0.
\label{eq:6}
\end{eqnarray}

Inequality Eq.~\ref{eq:6} is violated when two indistinguishable photons are injected into the device. Measured violations from injecting photons into 0, 1 and -1,+1 are plotted Fig.~\ref{Number of s violated}, with white data points representing no violation, and coloured data points representing the extent of violating Eq.~\ref{eq:6} for each pair of waveguides. This is quantified as a function of standard deviations $\sigma$ (computed from propagation of error from two photon coincidence detection, assuming Poissonian statistics), with the maximum violation reaching 76 standard deviations. Inequality Eq.~\ref{eq:6} is not violated when the photons of the input pairs are distinguishable. 

These demonstrations show uniquely non-classical behaviour of two identical particles, tunnelling through arrayed potential wells;  two photons initially prepared in a separable product state interfere in a generalisation of the Hong-Ou-Mandel effect \cite{ho-prl-59-2044} yielding non-classical spatial correlations. 
Increasing the photon number $n$ will emulate quantum walks on hyper-cubic graphs exponentially large in $n$, while exploiting quantum interference in three-dimensional directly written waveguides \cite{ma-oe-17-12546,ke-pra-81-023834} allows further increase in graph size \cite{ma-jpa-35-2745}. 
Enlarging the guided mode or decreasing the waveguide separation provides another increase in graph complexity. This requires a model beyond nearest neighbour coupling, tending towards a multimode interference slab waveguide when the channel separation goes to zero \cite{pe-mmi-in-prep}. Here, we have modelled and used arrays of waveguide with fixed propagation and coupling coefficients $\beta_{k} = \beta$ and $C_{k\pm1}=C$, but varying these parameters independently for each $k$ provides a means to engineer the quantum walk's precise graph structure. For example, varying these parameters randomly and independently allows investigation of correlated quantum walks in disordered systems and verify the effects of Anderson localisation \cite{la-arxiv-103-3657}, known to affect propagation of quantum information \cite{ke-pra-76-012315,la-prl-100-013906}. Reconfigurable waveguide circuits \cite{ob-natphot-3-687} will allow real-time control of the input state and the graph structure itself, enabling for example phase control over entangled ``NOON" input states to yield methods for simulating symmetric and anti-symmetric particles undergoing quantum walk \cite{br-prl-102-253904}.

\begin{small}
We thank N. Brunner, A. Laing and T. Rudolph for helpful discussion. This work was supported by EPSRC, ERC, QUANTIP, QIP IRC, IARPA, the Leverhulme Trust, and NSQI. J.L.OÕB. acknowledges a Royal Society Wolfson Merit Award.
\end{small}

\bibliography{JMatthewsQRWbib}

\clearpage
\section*{APPENDIX}

\begin{figure*}[b!]
\includegraphics[width=\textwidth]{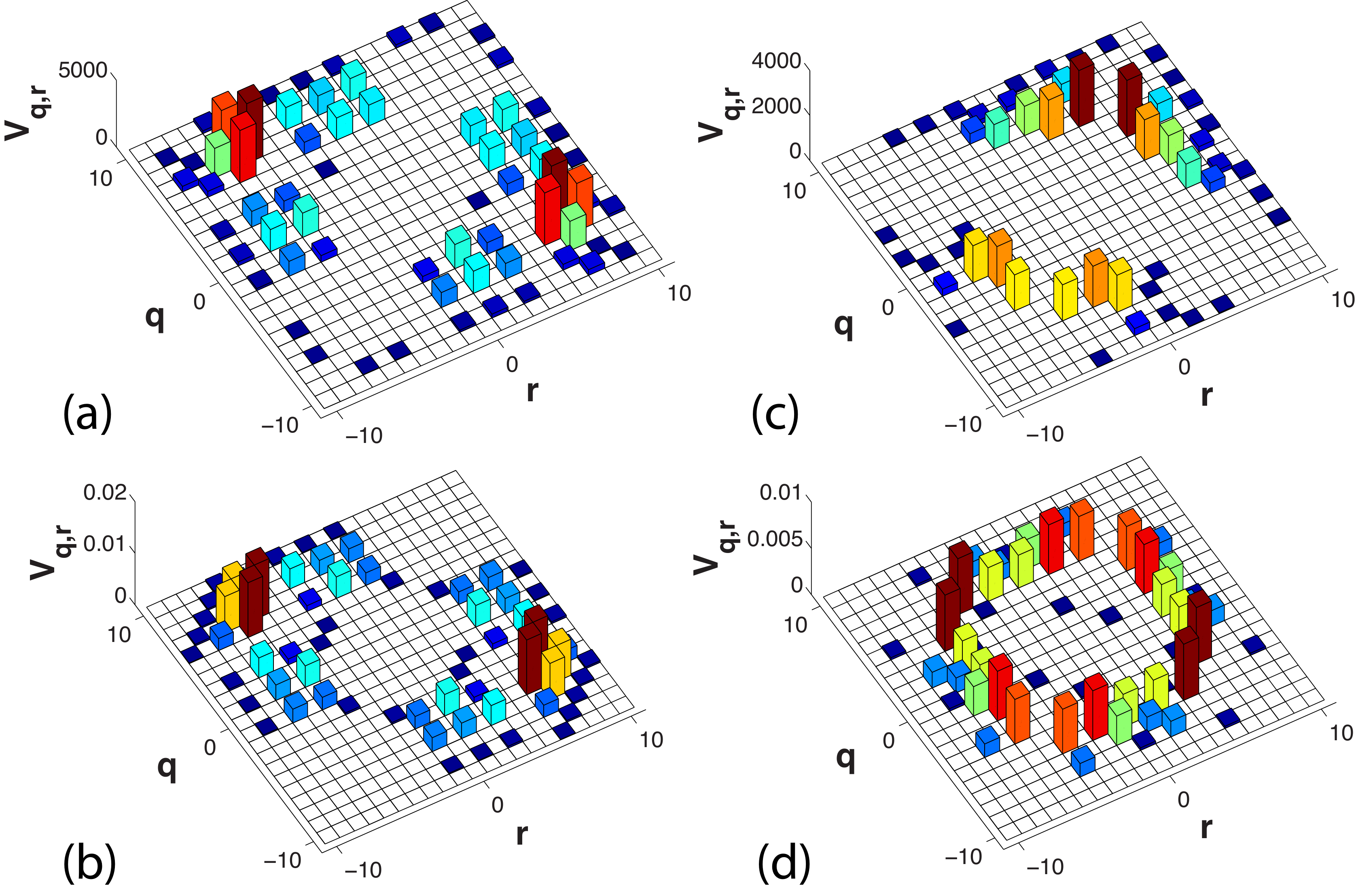}
\caption{Amount of violations of inequality Eq.~\ref{eq:6} of the main text. (a,b): measured and predicted for input state $ a_{0}^{\dagger}a_{+1}^{\dagger}\left|0\right\rangle $. (c,d): measured and predicted for input state $a_{-1}^{\dagger}a_{+1}^{\dagger}\left|0\right\rangle $.}
\label{ineq_viol}
\end{figure*}

\noindent\textbf{Devices:}
Each layer defining the photonic circuit was grown on thermally oxidised (8 $\mu$m) Silicon (Si $\langle100\rangle$) wafers with deposition performed in an Oxford parallel plate PECVD reactor utilising $SiH_4$ and $N_2O$ precursors. After annealing channel waveguides are obtained by standard lithography and reactive ion etching in $CHF_3 / O_2$ chemistry. The channel structures are covered by a PECVD silicon oxide ($SiO_2$) cladding layer. The core and cladding have a refractive index contrast of $\Delta = (n_{core}^2 - n_{cladding}^2)/2 n_{core}^2 = 4.4 \%$ enabling a waveguide bend radius of 600 $\mu$m resulting in compact 5mm long chips shown in Fig.~\ref{fig_chipPicture}(a) of the main text. The distance between waveguides in the coupled array (2.8 $\mu$m) defines the tunnelling rate of light between adjacent waveguides, while the waveguides at the end of the array spread out, at an equal rate of relative separation between adjacent waveguides, to a pitch 125 $\mu$m for coupling to optical fibre external to the chip; each pair of adjacent waveguides bend at the same rate until evanescent coupling is negligible, ensuring uniform coupling throughout the structure. The overall coupling efficiency through the chip is $\approx$10\% attributed to mode-mismatch between waveguides and input/output optical fibre. Injecting single photon input from SPDC is achieved using arrays of polarisation maintaining fibre, while collection uses arrays of multi-mode fibre. Device characterisation used $810$ nm light from a diode laser source launched into the central input with power intensity measured with a CCD camera; the resulting interference pattern is given in Fig.~\ref{fig_chipPicture}(c) and displays the ballistic dynamics equivalent to a single photon undergoing a continuous time quantum walk \cite{pe-prl-100-170506}. 
\\
\\
\noindent\textbf{Violations:}
Violation of the classical limit Eq.~\ref{eq:6} of the main text is quantified in Fig.~\ref{Number of s violated} by plotting the number of standard deviations measured violations reach beyond the classical limit. This violation is defined to be when the left hand side of Eq.~\ref{eq:6} is found to be negative:
\begin{eqnarray}
\Gamma_{q,r} - \frac{{2}}{3} \sqrt{\Gamma_{q,q},\Gamma_{r,r}}<0.
\end{eqnarray}
The measured value of this quantity for inputs $a^\dagger_0 a^\dagger_{+1}\ket{0}$ and $a^\dagger_{-1} a^\dagger_{+1}\ket{0}$ is given in figure  Fig.~\ref{ineq_viol} (a),(c), from which the number of standard deviations of Fig.~\ref{Number of s violated} are computed by propagating the Possonian standard deviation of measured two-photon statistics. The corresponding theoretical model of these violations are given in Fig.~\ref{ineq_viol} (b),(d).
\\
\\
\noindent\textbf{Dimension of multi-particle quantum walk Hilbert space:} In general Hilbert space representing $n$ distinguishable photons injected into a waveguide array of size $N$ is the tensor product of $n$ Hilbert spaces each of dimension $N$. This makes the combined state space $N^n$ dimensional. However, the photons still remain in a separable product state throughout evolution with unitary transform in the form of a tensor product of $n$ single photon unitary operators. These dynamics are trivial to reconstruct, requiring $n$ uses of the single photon evolution (equivalently bright laser light) for a uniform array, and $n$ separate evolutions in general.

For $n$ indistinguishable photons, the dimension of the Hilbert space also grows exponentially---bounded below by $N^n/n$. However, unlike the distinguishable case, evolution moves into non-separable states via non-classical interference---for example it is well-known that two-photon NOON states are generated by non-classical interference after a 50:50 beam splitter. For the case of two indistinguishable photons ($\hat{A} = a^\dagger_ja^\dagger_k$) injected to array of size $N$, the Hilbert space is of dimension $(N+1) N / 2$ (proven by induction). Similarly for 3-photons ($\hat{A} = a^\dagger_j a^\dagger_k a^\dagger_l$), the Hilbert space is of dimension $(N+2) (N+1) N / 6!$ (proven by induction); this emulates a CTQW on a three dimensional lattice. For $n$ indistinguishable photons non-classically interfering in a quantum walk, the dynamics cannot be decomposed into the dynamics of $n$ independent walks. The combination of distinctly non-classical states created through unitary evolution, and the dimension of the Hilbert space being exponentially large in $n$, implies simulation with non-interfering photons requires an exponential number of resources.

\end{document}